\documentstyle[12pt]{article}
\def\to{\rightarrow}

\def\NPB#1#2#3{{\em Nucl. Phys.} {\bf B#1} (19#2) #3}
\def\PLB#1#2#3{{\em Phys. Lett.} {\bf B#1} (19#2) #3}
\def\PLBold#1#2#3{{\em Phys. Lett.} {\bf#1B} (19#2) #3}
\def\PRD#1#2#3{{\em Phys. Rev.} {\bf D#1} (19#2) #3}
\def\PRL#1#2#3{{\em Phys. Rev. Lett.} {\bf#1} (19#2) #3}

\begin{document}
\begin{titlepage}
\begin{center}
\today     \hfill    IFT--P.047/97\\
~{} \hfill hep-ph/9708352\\

\vskip .1in

{\large \bf Triple Gauge Boson Production and Dynamical 
\\Symmetry Breaking at the Next Linear Collider}

\vskip 0.3in

Rogerio Rosenfeld\footnote{rosenfel@axp.ift.unesp.br} and
Alfonso R. Zerwekh\footnote{zerwekh@axp.ift.unesp.br}

\vskip 0.1in

{\em Instituto de F\'{\i}sica Te\'orica - Universidade Estadual Paulista\\
     Rua Pamplona, 145 - 01405--900 S\~ao Paulo - SP, Brazil}

\end{center}

\vskip .1in

\begin{abstract}
 We study in a model independent way the role of a techniomega
resonance in the  process  $e^+ e^- \to W^+ W^- Z$  at the 
 Next Linear Collider. 
\end{abstract}
\end{titlepage}

\newpage


The only sector of Standard Model that has not been directly tested
so far is the electroweak symmetry breaking sector. The usual Higgs potential 
with an elementary Higgs boson is not
satisfactory on the grounds of the triviality of a $\lambda \phi^4$ theory.
There are two alternatives to describe the symmetry breaking
sector that circumvent the triviality problem: supersymmetric models, with 
elementary Higgs bosons, and models with 
dynamical symmetry breaking, without elementary scalars.
In this work we will deal with the latter models \cite{dsb}. 

One of the places where the effects of dynamical symmetry breaking surely 
appear is in the
production of longitudinally polarized electroweak gauge bosons, since they
are directly related to the electroweak symmetry breaking of the Standard
Model \cite{equivalence}. Therefore, multiple gauge boson production may 
provide an important signature for these types of models.

The signature for large multiplicity ($\geq 7$)gauge boson production in 
hadron colliders 
was studied in ref. \cite{mpr} based on a scaling of pion 
multiplicity distribution at low energy electron-positron machines.
The gauge boson pair production process as a test of alternative models
of electroweak symmetry breaking has been analysed in detail 
recently in both hadron and
electron-positron colliders \cite{reviews}. 

The triple gauge boson production can occur through a techniomega resonance 
$\omega_T$, as first studied by
Rosenfeld and Rosner\cite{rr} for $\omega_T$ production from both vector 
meson dominance and gauge boson fusion processes in hadron colliders. 
However, backgrounds at hadron colliders are very severe \cite{gs}
and the rarer two body decay process $\omega_T \to \gamma (Z_T) + Z_L$,
where $(Z,W)_{L,T}$ denotes longitudinally and tranverselly polarized 
gauge bosons may be preferred, as shown by Chivukula and Golden \cite{cg}
in a minimal technicolor model. 

Recently, techniomega production has been studied in the context of 
multiscale technicolor models\cite{multi} in hadron colliders.
In multiscale walking 
technicolor models, where the $\omega_T$ can be as light as a few hundred GeV, 
there are pseudo-goldstone bosons ($\Pi_T$) that are not absorbed by the 
electroweak gauge
bosons and therefore remain as physical particles. In these models, the
decay mode $\omega_T \to \gamma \Pi_T \to \gamma \bar{b} b$ is the most
promising one in hadron colliders \cite{el}.

In this Letter we study the contribution of a techniomega resonance to the
production of three electroweak gauge bosons in the clean environment of
an electron-positron collider. We employ a model independent approach,
in the sense that we don't work in any particular model
but study discovery regions in a general parameter space without any 
theoretical prejudice. 
We then discuss the implications of our results to different specific
models.

The parameters that characterize the techniomega for our purposes are its 
mass $M_{\omega_T}$,
its total width $\Gamma_{\omega_T}$ and its partial width into 
$W_L^+ W_L^- Z_L$, $\Gamma_{WWZ}$. We make use of the equivalence
theorem \cite{equivalence} to relate the unphysical pseudo-goldstone bosons 
to the longitudinal components of the electroweak gauge bosons.

The coupling of the techniomega with a fermion-antifermion pair,
which is relevant for its production in $e^+ e^-$ colliders, can be estimated
by a generalized vector meson dominance 
mechanism which describes its mixing with the $B$ gauge boson of
$U(1)_Y$:
\begin{equation}
{\cal L}_{\omega_{T}B}=g_{\omega_{T}B}\omega^{\mu}B_{\mu}
\end{equation}
where the mixing constant $g_{\omega_{T}B}$ is given by:
\begin{equation}
g_{\omega_{T}B}= 2 \sqrt{2}\tan\theta_w m_w M_{\omega_T}(2A-1)
\end{equation}
where $m_w$ is the $W-$boson mass and  $A$ is the electric charge of the 
technifermion with weak isospin $+1/2$. 
In our calculations we use $A=2/3$. This coupling results in a partial width:
\begin{equation}
\Gamma(\omega_T  \to e^+ e^-) =  \frac{5\alpha m_{W}^{2} 
sin^{2}\theta_{w}}
{3m_{\omega_{T}} cos^{4}\theta_{w}}(2A-1)^2
\label{ee}
\end{equation}

The coupling constant describing the $\omega_T  W_L^+ W_L^- Z_L $ interaction
can be estimated from the partial width $\Gamma_{WWZ}$ using the 
effective interaction proposed in the context of QCD \cite{rudaz}:
\begin{equation}
{\cal L}_{\omega_{T}3\pi}=-ig_{\omega 3\pi}\epsilon^{\mu \nu \lambda \sigma}
\omega_{\mu}
\partial_{\nu} \pi^+  \partial_{\lambda} \pi^-  \partial_{\sigma} \pi^0
\end{equation}
which, using the equivalence theorem, results in:
\begin{equation}
\Gamma_{WWZ} = \frac{g_{\omega 3\pi}^2  M_{\omega_T}}{144 (2 \pi)^3} 
\int_{m_V}^{\frac{(M_{\omega_T}^2-3 m_V^2)}{2 M_{\omega_T}}} dE \;
\frac{(E^2-m_V^2)^{3/2} (M_{\omega_T}^2 - 2 M_{\omega_T} E -3 m_V^2)^{3/2}}
{( M_{\omega_T}^2 -  2 M_{\omega_T} E + m_V^2)^{1/2} } 
\label{width}
\end{equation}
where we used a value of $M_V = 85$ GeV.

We incorporated these new interactions  
into a HELAS-like \cite{helas} subroutine and 
with the help of the package MADGRAPH \cite{madgraph}, we computed
the process $e^+ e^- \to W^+ W^- Z$ in an extension of the Standard Model
containing the techniomega contribution. In this way we automatically
include the Standard Model irreducible background. Given the mass and the
partial width, we compute the relevant coupling constants via equations 
$2$ and $5$. 

We have checked our code by comparing its result for the partial widths
$\Gamma(\omega_T  \to e^+ e^-)$ and
$\Gamma_{WWZ}$ with the direct results from equations \ref{ee} and \ref{width}.
Another check was made by comparing the result for 
$e^+ e^- \to W_L^+ W_L^- Z_L$ obtained from the code without the Standard 
Model with a Breit-Wigner approximation near the peak:
\begin{equation}
\sigma=\frac{12\pi (s/m_{\omega_{T}}^2) \Gamma(\omega_{T} \rightarrow e^+ e^-)
\Gamma(\omega_{T} \rightarrow W^+ W^- Z^0)}{(s-m^{2}_{\omega_{T}})^2 + 
m^{2}_{\omega_{T}} \Gamma^{2}_{tot}}
\label{narrow}
\end{equation}


In Figure 1 we show three sets of curves representing the cross sections
for the process $e^+ e^- \to W^+ W^- Z$
for the masses $M_{\omega_T}=400, 500$ and $600$ GeV at 
 $\sqrt{s} = 500$ GeV as a function of the branching ratio
$BR(\omega_T \to W^+ W^- Z)$. Each set has three
curves, representing the different total widths $\Gamma_{\omega_T}=
20,\; 10$ and $5$ GeV. These are only representative values to illustrate our
results.   
We also show as horizontal dot-dashed lines the Standard Model cross section
as well as its $2 \sigma$ deviation assuming a luminosity of 
${\cal L} = 10$ fb$^{-1}$ and an efficiency for the reconstruction of the
three gauge bosons of $\epsilon = 12 \%$ \cite{eff}. No cuts are imposed.

For  $M_{\omega_T}$  near the center-of-mass energy, there is a  
sensitivity up to branching ratios as small
as $10^{-4}$ for a $2 \sigma$ deviation in the total cross section. Branching
ratios of the order of $10^{-2}$ can be reached for   $M_{\omega_T} = 400$ GeV
but for  $M_{\omega_T}= 600$ GeV one is sensitive only to large branching
ratios of order one.

The inversion of the order of the curves for the different sets of masses 
can be easily explained by inspection of the Breit-Wigner formula in
equation \ref{narrow}. At the resonance, the cross section is proportional to
$\Gamma_{\omega_T}^{-2}$, whereas  for the cases studied here the off-resonance
cross section is proportional to $\Gamma_{WWZ}$ which, for a fixed value of 
the branching ratio $BR(\omega_T \to W^+ W^- Z)$ is proportional to 
$\Gamma_{\omega_T}$.

In Figure 2 we show curves for an $e^+ e^-$ collider at $\sqrt{s} = 1000$ GeV,
for a total width of $\Gamma_{\omega_T}= 20$ GeV and techniomega masses at
$M_{\omega_T}= 950, 1000$ and $1050$ GeV. For masses at the center-of-mass 
energy, 
sensitivities up to $BR \simeq 0.5 \%$ can be achieved but the sensitivity
rapidly deteriorates for masses outside this range.

 Due to the fact that the signal is produced by an $s$-channel
resonance, its angular distribution is more central than the Standard Model
background, as can be seen in Figure 3, where the normalized gauge boson's 
angular distribution is shown for the case of $\sqrt{s}= 500$ GeV for the 
Standard Model and for the Standard Model plus a techniomega with mass
$M_{\omega_T} = 400$ GeV, total width  $\Gamma_{\omega_T}= 20$ GeV and partial
width  $\Gamma_{WWZ} = 0.4$ GeV. However, we found that a cut in the angular 
distribution of the $W^+$ or $W^-$ does not improve the significance due to 
the reduced rates.

The techniomega contributes only to the production of longitudinally
polarized gauge bosons and therefore one could enhance the signal by 
selecting that particular polarization in the final state. We demonstrate
this effect in Figure 4, where we show the cross section for  
 $e^+ e^- \to W^+ W^- Z_L$, where the $Z$ boson is longitudinal, at
$\sqrt{s}= 1000$ GeV, for $M_{\omega_T} = 950$ GeV and 
$\Gamma_{\omega_T}= 20$ GeV  as a function of the branching ratio
$BR(\omega_T \to W^+ W^- Z)$. The Standard Model cross section and its 
$2 \sigma$ deviation are shown as horizontal dot--dashed lines,  
assuming a luminosity of 
${\cal L} = 10$ fb$^{-1}$ and the same efficiency for the reconstruction of 
the three gauge bosons, $\epsilon = 12 \%$. This is to be compared with the 
dotted line in Figure 2. We note an increase in the sensitivity of the cross 
section to the presence of a techniomega but due to the reduced statistics
no major improvement seems to be achieved.

We conclude by commenting on some specific models. In usual technicolor 
models the techniomega mass and width are estimated by a simple scaling
of QCD, resulting in typically $M_{\omega_T} \simeq 2$ TeV and 
$\Gamma_{\omega_T} \simeq 100$ GeV.
However, these simple models run into problems with generating large fermion
masses while keeping flavor changing neutral currents at acceptable levels.

Multiscale models \cite{multi}, that appear naturally in walking technicolor
\cite{walk} and topcolor-assisted technicolor\cite{tc} models,  can ameliorate
this problem at the same time predicting
vector resonances with lower masses than the simple scaled up models. 
These models have
many extra technipions $\Pi_T$'s that are not absorbed by the massless 
gauge bosons.

The techniomega would preferably decay into three $\Pi_T$'s:
\begin{equation}
\frac{\Gamma (\omega_T \to \Pi_T^+ \Pi_T^- \Pi_T^0)}{\Gamma (\omega_T \to
W_L^+ W_L^- Z_L)} \propto \frac{1}{\sin^6 \chi}
\end{equation}
where $\chi$ is a mixing angle related to the ratio of the two different 
energy scales in a particular model. 

It may be possible that the three  $\Pi_T$'s channel is closed due to 
technipion mass enhancements in some of these models. In this case
one would expect branching ratios of the order of:
\begin{equation}
BR(\omega_T \to W_L^+ W_L^- Z_L) \simeq \sin^4 \chi = 
\left\{ \begin{array}{ll}
1.2 \% & \mbox{for $\sin \chi = 1/3$} \\
6.2 \% & \mbox{for $\sin \chi = 1/2$}
	\end{array}  \right.
\end{equation}
which falls in the sensitive region for some of the cases studied here.

In summary, we have studied the role of a techniomega resonance in the  
production of three gauge bosons in $e^+ e^-$ colliders.
We considered the effects on the total cross section, the angular distribution,
and in the total cross section for a longitudinally polarized $Z$ boson
in the final state.
Due to the narrow bandwidth of the machine, one is sensitive to techniomega
masses close to the center-of-mass energy. In that case, one can be sensitive
to rather small branching ratios. Initial state radiation may not
be able to improve our results significantly.
If the techniomega resonance is found at a hadron machine in the mode
$\omega_T \to \gamma \pi_T \to \gamma \bar{b} b$ \cite{el}, its properties 
can be further studied at an $e^+ e^-$ collider running at its mass,
which could also provide useful information about the underlying technicolor
model.

\vspace{2cm}

This work was supported by Conselho Nacional de Desenvolvimento 
Cient\'{\i}fico e Tecnol\'ogico (CNPq) and Funda\c{c}\~ao de Amparo \`a 
Pesquisa do Estado de S\~ao Paulo (FAPESP). We thank Gustavo Burdman and 
Jonathan Rosner for a critical reading of the paper.
RR gratefully
acknowledges the hospitality at the Fermilab Theory Group, where this work
was completed.

\newpage

{\Large \bf Figure Caption}

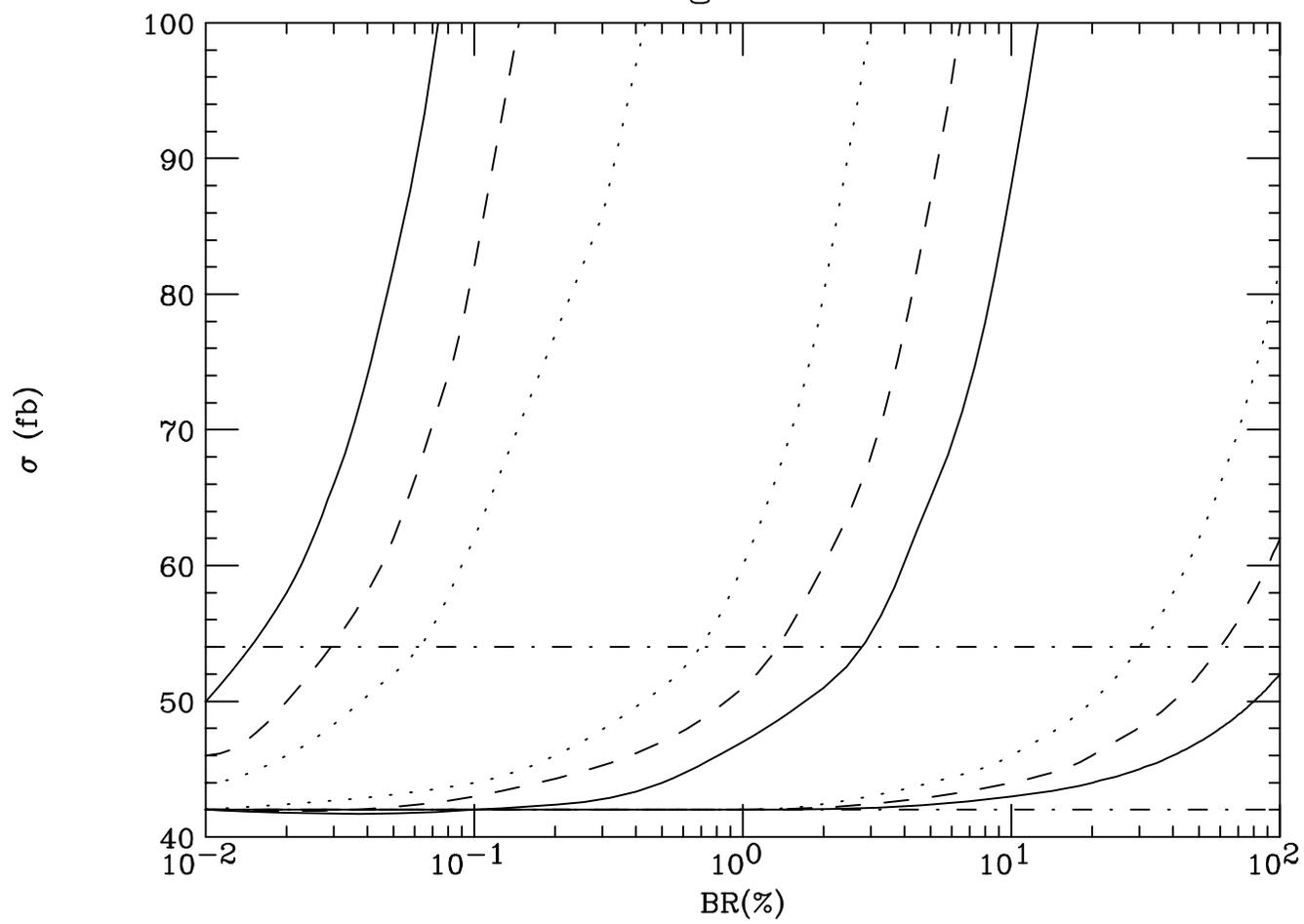
{\bf Figure 1:}\\
Cross section for  $e^+ e^- \to W^+ W^- Z$ at $\sqrt{s} = 500$ GeV as a 
function of the branching ratio $BR(\omega_T \to W^+ W^- Z)$ without 
any cuts.
The  three sets of curves are the cross sections
for the process $e^+ e^- \to W^+ W^- Z$
for the masses $M_{\omega_T}=400$ GeV (middle set), $500$ GeV (left set) and 
$600$ GeV (right set).
 Each set has three
curves, representing the different total widths $\Gamma_{\omega_T}=
20$ GeV (dotted line), $10$ GeV (dashed line) and $5$ GeV (solid line). 
The Standard Model cross section
as well as its $2 \sigma$ deviation assuming a luminosity of 
${\cal L} = 10$ fb$^{-1}$ and an efficiency for the reconstruction of the
three gauge bosons of $\epsilon = 12 \%$ are shown as horizontal
dot-dashed lines.

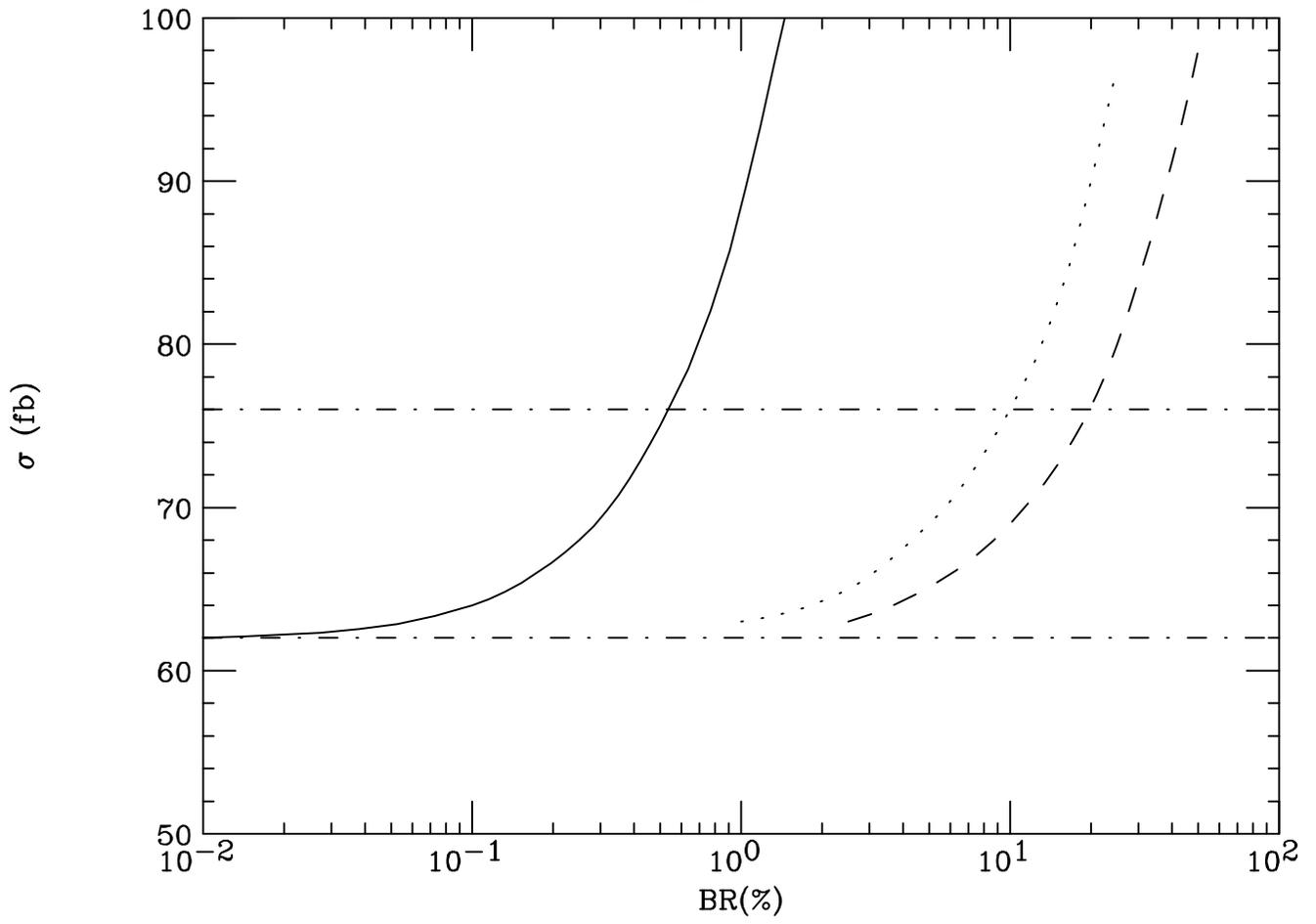
{\bf Figure 2:} \\
Same as in Figure 1 but for $\sqrt{s} = 1000$ GeV and  a fixed total width  
$\Gamma_{\omega_T}=20$ GeV. The different techniomega masses are 
$M_{\omega_T}=950$ GeV (dotted line), $1000$ GeV (solid line) and 
$1050$ GeV (dashed line).

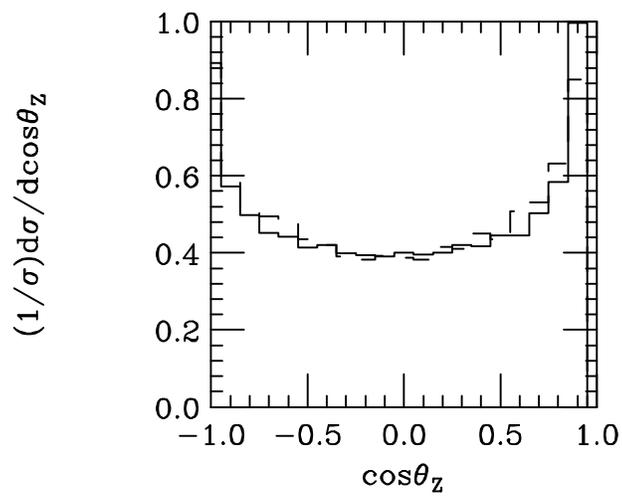
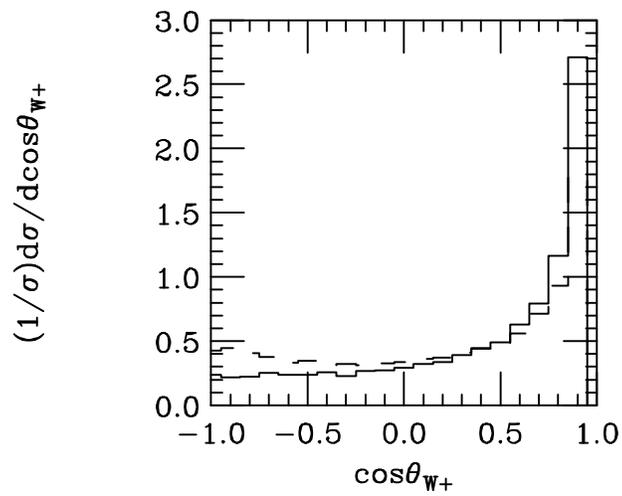
{\bf Figure 3:} \\
Normalized angular distribution for the three final state gauge bosons
for $e^+ e^- \to W^+ W^- Z$ at $\sqrt{s} = 500$ GeV.
Solid line is the Standard Model result and dashed line is the Standard 
Model plus a techniomega with $M_{\omega_T}=400$ GeV,  $\Gamma_{\omega_T}=
20$ GeV and $\Gamma_{WWZ} = 0.4$ GeV.

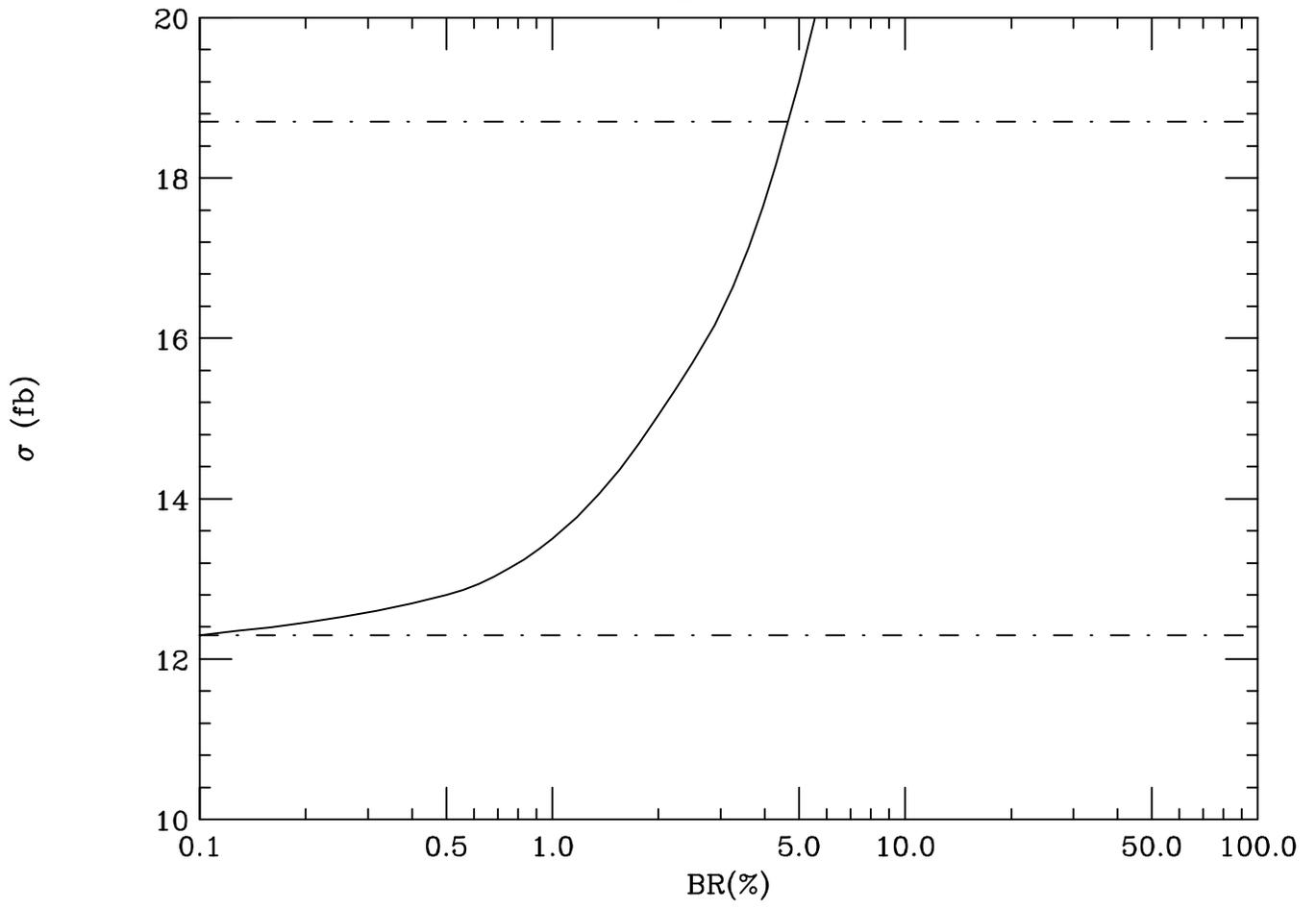
{\bf Figure 4:}\\ 
Same as the dotted line in Figure 2 but requiring a longitudinally polarized
$Z$ boson in the final state. The Standard Model cross section
as well as its $2 \sigma$ deviation assuming a luminosity of 
${\cal L} = 10$ fb$^{-1}$ and an efficiency for the reconstruction of the
three gauge bosons of $\epsilon = 12 \%$ are shown as horizontal
dot-dashed lines.

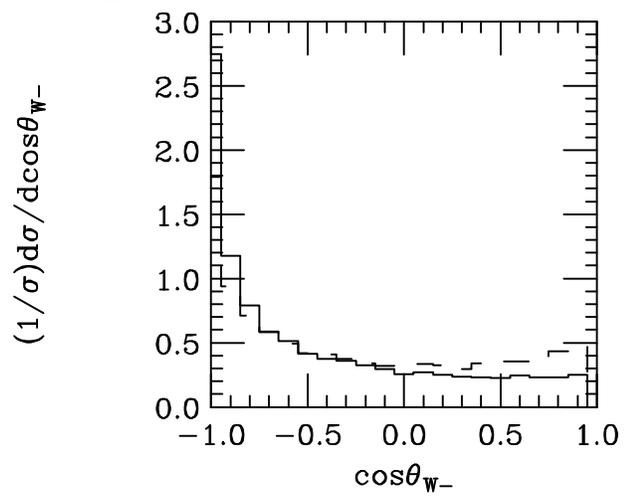
\end{document}